\documentstyle[12pt]{article}
\newcommand{\doublespace}{\renewcommand{\baselinestretch}{1.75}
\Large\normalsize}
\begin{document}
\doublespace
\begin{titlepage}

\centerline{\bf ON THE DISTRIBUTION OF GRAVITATIONAL ENERGY} 
\centerline{\bf IN THE DE SITTER SPACE}
\vskip 2.0cm
\centerline{Jos\'e W. Maluf$^*$}
\centerline{Universidade de Bras\'ilia}
\centerline{C.P. 04385}
\centerline{70.919-970 Bras\'ilia DF}
\centerline{Brazil}
\bigskip

\date{}

\begin{abstract}
We calculate the total gravitational energy and the
gravitational energy density of the de Sitter space using
the definition of localized gravitational energy that naturally 
arises in the framework of the teleparallel equivalent of general
relativity. We find that the gravitational energy can only be
defined within the cosmological horizon and
is largely concentrated in regions far from the center 
of spherical symmetry, i.e.,
in the vicinity of the maximal spacelike radial coordinate
$R=\sqrt{3\over \Lambda}$. The smaller the cosmological
constant, the farther the concentration of energy. 
This result complies with the phenomenological features
of the de Sitter solution, namely, the existence of a
radial acceleration directed away from the center of symmetry
experienced by a test particle in the de Sitter space.
Einstein already contemplated the de Sitter 
solution as a world with a surface distribution of matter, 
a picture which is in agreement with the present  analysis.

\end{abstract}
\thispagestyle{empty}
\vfill
\par
\noindent (*) e-mail: wadih@guarany.cpd.unb.br
\end{titlepage}
\newpage

\noindent {\bf I. Introduction}\par
\bigskip
\noindent The difficulty in defining gravitational energy density
within the framework of the Hilbert-Einstein Lagrangian formulation
has led to the belief that the gravitational energy cannot be 
localized. It is widely assumed that an expression for the 
localized energy density of the gravitational field does not exist.
However it is well known that the total energy of a given
asymptotically flat spacetime can be calculated by means of
pseudotensor methods, which make use of coordinate dependent 
expressions. A different approach to the construction of an energy
expression for the gravitational field is based on the idea of
quasilocal energy. The quasilocal definition of energy, momentum
and angular momentum associates these quantities to an arbitrary
spacelike two surface $S$ in an arbitrary spacetime manifold.
The inexistence of an unequivocal
definiton of gravitational energy still
remains an actual problem, important in its own 
right. Furthermore such definiton may play a major 
role in the thermodynamics of self-gravitating systems. This problem
has been recently addressed in ref.\cite{Brown}, where a 
comprehensive bibliography on quasi-local energy is 
presented. Although all attempts so far have led to interesting
mathematical developments, they did not allow the achievement of
a definite solution, either because of conceptual or mathematical
difficulties.

Recently the problem of localization of energy in general relativity
has been reconsidered in the framework of the teleparallel equivalent 
of general relativity (TEGR)\cite{Maluf1}.  The Lagrangian 
formulation of the TEGR is established by means of the tetrad field
$e^a\,_\mu$ and the spin affine connection $\omega_{\mu ab}$,
which are taken to be completely independent field variables,
even at the level of field equations.
This formulation has been investigated in the past in
the context of Poincar\'e gauge theories\cite{Hehl1,Hehl2}. However,
as we will explain ahead, this is not an alternative theory of gravity.
This is just an {\it alternative formulation} of general relativity, 
in which the curvature tensor constructed out of $\omega_{\mu ab}$
vanishes, but the torsion tensor is non-vanishing.  
The physical content of the theory is dictated by Einstein's
equations. As we will show, in this alternative geometrical 
formulation the gravitational energy density can be naturally defined.

The expression for the localized 
energy density of the gravitational field has arisen in the 
context of the Hamiltonian formulation of the TEGR\cite{Maluf2}. 
It has been demonstrated
that under a suitable gauge fixing of $\omega_{\mu ab}$, already
at the Lagrangian level, the Hamiltonian formulation of the TEGR 
is well defined\cite{Maluf2}. The resulting constraints are first 
class  constraints. In fact  the Hamiltonian formulation
looks very much similar to the 
the usual ADM formulation\cite{ADM}. However there are crucial
differences. The integral form of the 
Hamiltonian constraint equation $C=0$ in the TEGR can be written 
in  the form $C=H-E_{ADM}=0$, when we restrict considerations
to asymptotically flat spacetimes\cite{Maluf1}.
The quantity $\varepsilon(x)$
which appears in the expression of $C$ and
which under intergration yields $E_{ADM}$ is recognized as the 
gravitational energy density. We have applied the expression of
$\varepsilon(x)$ to the calculation of the energy inside a surface
of constant radius, both for the Schwarzschild\cite{Maluf1} and for
the Kerr metric\cite{Maluf3}, and the results are remarkably the 
same as those obtained by means of the quasilocal energy definition
proposed by
Brown and York\cite{Brown}. Moreover, the calculational scheme is
rather simple, as we will see shortly, and is exempt of some
complications inherent to the latter.
Therefore for asymptotically flat
spacetimes the gravitational energy density has a definite 
and unambiguous expression within the framework of the TEGR.

We recall that the gravitational energy can also be calculated
by means of the surface term that appears in the expression
of the gravitational Hamiltonian\cite{Regge,Faddeev}. However,
such surface term yields only the {\it total} gravitational 
energy, as the integration has to be necessarily carried out 
over the whole three dimensional spacelike hypersurface, in
which case the lapse function $N(x)$ goes over into its 
asymptotic value $N \rightarrow 1$ at spatial infinity.

The action integrals for spacetimes with different topologies
require surface terms that are specific to each topology. 
Thus the corresponding Hamiltonian also acquires a surface term
that is determined by the topological boundary 
conditions\cite{Hawking}. However the Hamiltonian constraint
for a spacetime foliated by spacelike hypersurfaces
always has the same basic structure, irrespective of boundary  
conditions ({\it additional} terms such as the cosmological
constant may appear in the Hamiltonian
constraint, as we will see ahead in eq.(10) ).

Therefore the question immediately arises as to whether the 
Hamiltonian constraint equation in the TEGR can always be 
writtem as  $C=H-E=0$, in which case $\varepsilon(x)$ would 
be the gravitational energy density for any curved spacetime.

One of the simplest deviations from asymptotically flat geometries
are spacetimes with conical defects.
We have applied our expression of gravitational energy density
to the calculation of the energy per unit length of defects
of topological nature, which include disclinations, i.e.,
cosmic strings, and dislocations\cite{Maluf4}. The result is
quite encouraging. We arrive at precisely the same well known 
expression for the energy per unit length of a cosmic string (not
even multiplicative factors have to be adjusted).
Moreover the total energy of a dislocation is zero, a result
which is in close analogy with the statements of the theory of
elasticity, which asserts that disclinations and dislocations 
are defects which require high energy and low energy, respectively.

In this paper we consider the de Sitter space, which is 
topologically of the type $S^3\times R$. We restrict the 
considerations to the static region within the cosmological
horizon (i.e., the region for which  
$-g_{00}>0$ )  and calculate both the
total energy and the distribution of energy along the radial 
direction. Again the result is rather remarkable. We will show 
that the cosmological constant induces a distribution of 
gravitational energy in such a way that the energy is largely
concentrated in the peripheral region, i.e., in the vicinity of 
the maximal spacelike radial coordinate $R=\sqrt{3\over \Lambda}$.
As we will show in
section III, this picture is in total agreement with the 
phenomenological features of the de Sitter solution, and is 
as well in agreement with Einstein's belief, according to which 
the de Sitter's solution represents a spacetime with a surface 
distribution of matter\cite{Pauli}. This fact strongly supports
the validity of our expression for the gravitational 
energy density and also represents a clear indication
that the Hamiltonian constraint equation in the TEGR can be 
unambiguously interpreted as an energy equation of the type
$H-E=0$.                

\bigskip
\noindent Notation: spacetime indices $\mu, \nu, ...$ 
and local Lorentz indices
$a, b, ...$ run from 0 to 3. In the 3+1 decomposition latin indices 
from the middle of the alphabet indicate space indices according to
$\mu=0,i,\;\;a=(0),(i)$. The tetrad field $e^a\,_\mu$ and
the spin connection $\omega_{\mu ab}$ yield the usual definitions
of the torsion and curvature tensors:  $R^a\,_{b \mu \nu}=
\partial_\mu \omega_\nu\,^a\,_b +
\omega_\mu\,^a\,_c\omega_\nu\,^c\,_b\,-\,...$,
$T^a\,_{\mu \nu}=\partial_\mu e^a\,_\nu+
\omega_\mu\,^a\,_b\,e^b\,_\nu\,-\,...$. The flat spacetime metric 
is fixed by $\eta_{(0)(0)}=-1$. \\


\bigskip
\bigskip
\noindent {\bf II. The Lagrangian and Hamiltonian formulations
of the TEGR}\par

\bigskip

In the TEGR the tetrad field $e^a\,_\mu$ and the spin connection
$\omega_{\mu ab}$ are independent field variables, not related
by any of the field equations. The  spin connection
is enforced to satisfy the condition of zero curvature. 
The Lagrangian density in empty spacetime is given 
by\cite{Maluf1,Maluf2}

$$L(e,\omega,\lambda)\;=\;-ke({1\over 4}T^{abc}T_{abc}\,+\,
{1\over 2}T^{abc}T_{bac}\,-\,T^aT_a)\;+\;
e\lambda^{ab\mu\nu}R_{ab\mu\nu}(\omega)\;.\eqno(1)$$

\noindent where $k={1\over {16\pi G}}$, $G$ is the gravitational 
constant; $e\,=\,det(e^a\,_\mu)$, $\lambda^{ab\mu\nu}$ are 
Lagrange multipliers and $T_a$ is the trace of the torsion tensor
defined by $T_a=T^b\,_{ba}$.   

The equivalence of the TEGR with Einstein's general relativity is         
guaranteed by the identity

$$eR(e,\omega)\;=\;eR(e)\,+\,
e({1\over 4}T^{abc}T_{abc}\,
+\,{1\over 2}T^{abc}T_{acb}\,-\,T^aT_a)\,-\,
2\partial_\mu(eT^{\mu})\;,\eqno(2)$$

\noindent which is obtained by just substituting the arbitrary
spin connection $\omega_{\mu ab}\,=\,^o\omega_{\mu ab}(e)\,+\,
K_{\mu ab}$ in the scalar curvature tensor $R(e,\omega)$ in the
left hand side of (2); $^o\omega_{\mu ab}(e)$ is the Levi-Civita 
connection and $K_{\mu ab}\,=\,
{1\over 2}e_a\,^\lambda e_b\,^\nu(T_{\lambda \mu \nu}+
T_{\nu \lambda \mu}-T_{\mu \nu \lambda})$ is the contorsion tensor.
The vanishing of $R^a\,_{b\mu\nu}(\omega)$, which is one of the
field equations derived from (1), implies the equivalence of 
the scalar curvature $R(e)$, constructed out of $e^a\,_\mu$ only, 
and the quadratic combination of the torsion tensor. It also
ensures that the field equation arising from the variation of
$L$ with respect to $e^a\,_\mu$ is strictly equivalent to
Einstein's equations in tetrad form. Let  
${{\delta L}\over{\delta e^{a\mu}}}=0$ denote the field equation 
satisfied by $e_{a\mu}$. It can be shown by explicit calculations
that

$${{\delta L}\over{\delta e^{a\mu}}}\;=\;{1\over 2}\lbrace R_{a\mu}-
{1\over 2}e_{a\mu}R(e)\rbrace\;.$$

\noindent (we refer the reader to
ref.\cite{Maluf2} for additional details).

For {\it asymptoticaly flat} spacetimes the total divergence 
in (2) does {\it not} contribute to the action integral.  Therefore 
the latter does not require additional surface terms, as it is 
already invariant under coordinate transformations that preserve 
the asymptotic structure of the field quantities\cite{Faddeev}. 
It is well known that for compact geometries a surface term has
to be included in the action, in order to make the variations of
the field variables well defined. This surface term is constructed 
out of the trace of the extrinsic curvature on the boundary. 
However we will no longer worry about surface terms in the 
Lagrangian or in the Hamiltonian, as we will be interested only
in the constraint structure of the theory.

The Hamiltonian formulation of the TEGR can be successfully 
implemented if we fix the gauge $\omega_{0ab}=0$ from the 
outset, since in this case the constraints 
constitute a {\it first class} set\cite{Maluf2}.
The condition $\omega_{0ab}=0$ is achieved by breaking the local
Lorentz symmetry of (1). We still make use of the residual time
independent gauge symmetry to fix the usual time gauge condition
$e_{(k)}\,^0\,=\,e_{(0)i}\,=\,0$. Because of $\omega_{0ab}=0$,
$H$ does not depend on $P^{kab}$, the momentum canonically 
conjugated to $\omega_{kab}$. Therefore arbitrary variations of
$L=p\dot q -H$ with respect to $P^{kab}$ yields 
$\dot \omega_{kab}=0$. Thus in view of $\omega_{0ab}=0$, 
$\omega_{kab}$ drops out from our considerations. The above 
gauge fixing can be understood as the fixation of a {\it global}
reference frame.    

Under the above gauge fixing the canonical action integral obtained
from (1) becomes\cite{Maluf2}

$$A_{TL}\;=\;\int d^4x\lbrace \Pi^{(j)k}\dot e_{(j)k}\,-\,H\rbrace\;,
\eqno(3)$$

$$H\;=\;NC\,+\,N^iC_i\,+\,\Sigma_{mn}\Pi^{mn}\;.\eqno(4)$$

\noindent In expression (4) above we are omitting surface terms.
$N$ and $N^i$ are the lapse and shift functions, 
$\Pi^{mn}=e_{(j)}\,^m\Pi^{(j)n}$  and 
$\Sigma_{mn}=-\Sigma_{nm}$ are Lagrange multipliers. The constraints
are defined by 

$$ C\;=\;\partial_j(2keT^j)\,-\,ke\Sigma^{kij}T_{kij}\,-\,
{1\over {4ke}}(\Pi^{ij}\Pi_{ji}-{1\over 2}\Pi^2)\;,\eqno(5)$$

$$C_k\;=\;-e_{(j)k}\partial_i\Pi^{(j)i}\,-\,
\Pi^{(j)i}T_{(j)ik}\;,\eqno(6)$$

\noindent with $e=det(e_{(j)k})$ and $T^i\,=\,g^{ik}e^{(j)l}T_{(j)lk}$, 
$\;T_{(j)lk}=\partial_l e_{(j)k}-\partial_k e_{(j)l}$.
We remark that (3) and (4) are invariant under {\it global} SO(3) and
general coordinate transformations (in eqs. (1) and (2) $e$ is
the determinant of the {\it spacetime} tetrad field; from eq.(3)
on $e$ stands for  the determinant of the triads restricted to the 
three dimensional spacelike hypersurface).

If we assume 
the asymptotic behaviour $e_{(j)k}\approx \eta_{jk}+
{1\over 2}h_{jk}({1\over r})$ for $r \rightarrow \infty$, which is 
appropriate for an asymptotically flat spacetime, then in view 
of the relation

$${1\over {8\pi G}}\int d^3x\partial_j(eT^j)\;=\;
{1\over {16\pi G}}\int_S dS_k(\partial_ih_{ik}-\partial_kh_{ii})
\; \equiv \; E_{ADM}\;\eqno(7)$$

\noindent where the surface integral is evaluated for 
$r \rightarrow \infty$, we note that the integral form of 
the Hamiltonian constraint $C=0$ may be rewritten as

$$\int d^3x\biggl\{ ke\Sigma^{kij}T_{kij}+
{1\over {4ke}}(\Pi^{ij}\Pi_{ji}-{1\over 2}\Pi^2)\biggr\}
\;=\;E_{ADM}\;.\eqno(8)$$

\noindent The integration is over the whole three dimensional
space. Given that $\partial_j(eT^j)$ is a scalar  density,
from (7) and (8) we define the gravitational
energy density enclosed by a volume V of the space as\cite{Maluf1}

$$E_g\;=\;{1\over {8\pi G}}\int_V d^3x\partial_j(eT^j)\;.\eqno(9)$$  

\noindent It must be noted that this expression is 
also invariant under global SO(3) transformations. 

One is immediately led to ask whether the Hamiltonian constraint
for topologically different spacetimes can also be written
as eq.(8). In the next section we will consider the de Sitter
space. Before addressing the latter, let us recall here 
some applications of $E_g$. We have calculated the gravitational
energy inside a surface of constant radius $r_o$ both for the
Schwarzschild\cite{Maluf1} and for the Kerr solution\cite{Maluf3},
using Boyer and Lindquist coordinates\cite{Kerr,Boyer}.
These quantities have also been calculated by means of Brown and
York's precedure, in refs.\cite{Brown} and
\cite{Martinez}, respectively. The expressions found by  using 
(9) are in total agreement with those obtained via the method of 
ref.\cite{Brown}. Moreover $E_g$ can be calculated for any volume
in the three-dimensional spacelike hypersurface, as least through
numerical integration, whereas the evaluation of the energy
in ref.\cite{Martinez} can only be carried out in the limit of
slow rotation of the black hole 
(the application of Brown and York's procedure to the Kerr 
solution with arbitrary parameters meets some technical
difficulties, as discussed in ref.\cite{Martinez}).

Definition (9) has also been applied to a class of conical
spacetime defects, in which disclinations (cosmic strings) 
and dislocations are considered altogether.
For the spacetime of a single cosmic
string, i.e., for a pure disclination, 
we obtain precisely the well known value of energy per
unit length of the string\cite{Maluf4}. Furthermore the {\it total}
gravitational energy for a pure dislocation vanishes. This is a
very interesting result, because we know from the theory of
elasticity that disclinations are defects that require 
a large ammount of energy to be formed, whereas dislocations
require low energy (see sections 6.3.2 and 6.5 of ref.\cite{Rivier}
for a discussion as to why the energy demanded for the formation
of a disclination in a crystal is very high).
Therefore  the above calculations of energy are in 
close agreement with the statements of the theory of elasticity
(in this respect we recall that attempts were made long time ago 
which envisaged the spacetime as a continuum with microstructure
(see\cite{Hehl2}, section 1.2)). \\

\bigskip
\bigskip

\noindent {\bf III. Gravitational Energy in the de Sitter Space}\par
\bigskip

\noindent We will consider now the theory defined by the 
Lagrangian density (1) supplemented by a term containing the 
cosmological constant $\Lambda$. Thus we add to (1) the quantity
$2\,^4e\Lambda$, where $^4e=Ne$ is the determinant of the
{\it spacetime} tetrad field $e_{a\mu}$. This additional term 
will contribute to the action integral (3) only as an extra 
term of the Hamiltonian constraint. The new Hamiltonian
constraint reads

$$ C\;=\;\partial_j(2keT^j)\,-\,ke\Sigma^{kij}T_{kij}\,-\,
{1\over {4ke}}(\Pi^{ij}\Pi_{ji}-{1\over 2}\Pi^2)
\,-2e\Lambda \;,\eqno(10)$$

\noindent The most general spherically symmetric solution of the
field equations with a positive cosmological constant is the 
Schwarzschild-de Sitter solution (throughout this section we
will make $G=1$):

$$ds^2\;=\;-\biggl(1-{{2m}\over r}-{{r^2}\over {R^2}}\biggr)
dt^2\,+\,\biggl(1-{{2m}\over r}-{{r^2}\over {R^2}}\biggr)^{-1} 
dr^2\,+\,r^2\,d\theta^2\,+\,r^2\,sin^2\theta\,d\phi^2\;.\eqno(11)$$

\noindent This metric represents the gravitational field of a 
particle of mass $m$ located at the origin of a globally hyperbolic 
spacetime. The vacuum 
solution, obtained by setting $m=0$ in (11), is the de Sitter 
solution. $R$ is the maximal spacelike radial coordinate 
for the (vacuum) de Sitter space
and is given by  $R=\sqrt{{3\over \Lambda}}$. 

Strictly speaking de Sitter spacetime corresponds to a 
four-dimensional surface in a flat five-dimensional space with 
metric (-,+,+,+,+) described by

$$-z_0^2\,+\,z_1^2\,+\,z_2^2\,+\,z_3^2\,+\,z_3^2\,+\,z_4^2\;=\;
{3\over \Lambda}\;,\;\;\;\Lambda>0\;.$$

\noindent The coordinates $(t,r,,\theta, \phi)$ in (11) cover only
half of the space defined by the relation above. However we will
be interested just in (11), as it suffices for our purposes.
Moreover we will restrict the considerations to the physical
region between the Schwarzschild (black hole) and the 
cosmological horizons.

Expression (9) allows us to calculate the gravitational energy
for any volume in space.
We wish to obtain the energy contained within a surface of constant
radius $r_o$. For this purpose  we will calculate
$eT^1\,=\,eT^r$ for a spacetime whose spacelike  section is
described by the line element

$$dl^2\;=\;\alpha^2\,dr^2\,+\,r^2\,d\theta^2\,+\,
r^2\,sin^2\theta\,d\phi^2\;.\eqno(12)$$

\noindent where $\alpha$ is a function of the coordinate $r$. The
triads that correspond to (12) are given by

$$e_{(k)i}\;=\;\pmatrix{ \alpha\,sin\theta cos\phi&
r\,cos\theta cos\phi& -r\,sin\theta sin\phi\cr
\alpha\, sin\theta sin\phi&
r\,cos\theta sin\phi& r\,sin\theta cos\phi\cr
\alpha\,cos\theta& -r\,sin\theta& 0}\;.\eqno(13)$$

\noindent $(k)$ is the line index, and $i$ is the column index. 

The determinant $e$ of (13) reads $e=\alpha \,r^2sin\theta$.
After a lengthy but otherwise straightforward calculation of 

$$eT^1\;=\;e\,g^{1j}\,g^{im}\,e^{(k)}\,_mT_{(k)ij}$$

\noindent we arrive at

$$eT^1\;=\;2r\,sin\theta \biggl( 1-{1\over{\alpha}}\biggr)\;.
\eqno(14)$$

\noindent Therefore for the Schwarzschild-de Sitter solution we
have

$$eT^1\;=\; 2r\,sin\theta\biggl(1-\sqrt{1-{{2m}\over r}-
{{r^2}\over {R^2}}}\biggr)\;.\eqno(15)$$

\noindent The energy contained within a surface of constant
radius $r_o$ is thus given by

$$E_g\;=\;{1\over {8\pi}}\int_S d\theta\,d\phi\,eT^1\;=\;
r_o\biggl( 1-\sqrt{ 1-{{2m}\over {r_o}}-{{r_o^2}\over {R^2}}}
\biggr)\;,\eqno(16)$$

\noindent where $S$ is a surface of constant radius $r_o$.

Let us evaluate expression (16) for the range of values of
$r_o$ such that 

$${{2m}\over {r_o}}\ll 1\;\;,\;\;{{r_o}^2\over {R^2}}\ll 1\;,$$

\noindent in which case we  assume the cosmological constant
to be very small. Expanding (16) and neglecting all powers of
both ${{2m}\over {r_o}}$ and ${{r_o}^2\over {R^2}}$ 
we arrive at

$$E_g\;=r_o\;+\;m\;\;\equiv\; \;E_{dS}\;+\;m\;.\eqno(17)$$

\noindent $E_{dS}$ is the value of energy we would obtain in 
the absence of the mass $m$, and therefore it represents the 
background (vacuum) energy. Upon subtraction of this term we
obtain the standard ADM value of energy for a particle of mass
$m$. Of course in (17) we expect $r_o$ to be much larger than
$m$.

The total gravitational energy contained within the cosmological
horizon can be easily calculated, but for this purpose one 
has to find the roots of the equation  
$1-{{2m}\over r}-{{r^2}\over {R^2}}\,=0$. The result is not 
illuminating. If $R\gg m$ we find that $E_g^{total}=r_{max}$, 
where $r_{max}$ is slightly smaller than $R$ and is a solution 
of the equation above.
In what follows we will rather consider the vacuum 
de Sitter solution only, since in this case the analysis of the
energy density is most easily carried out, and the main features
are not altered by the introduction of a mass $m$ at $r=0$.

Before proceeding we mention that the present analysis is 
different from that carried out by Abbott and Deser\cite{Abbott}. 
These authors provide an expression for the energy of the 
gravitational field {\it about} the de Sitter background, i.e.,
they calculate the energy of a field configuration that deviates
from the de Sitter metric and which vanish at infinity.
In contrast, by means of expression (9) we can compute
the energy of the whole gravitational field configuration,
including the background.

The total gravitational energy $E_{dS}$ contained in the 
physical region of the vacuum 
de Sitter space is obtained from (16) by making $m=0$ and
$r_o=R$:

$$E_{dS}\;=\;R\;=\;\sqrt{3\over \Lambda}\;.\eqno(18)$$

\noindent The total volume of the compact spacelike section
equals  $2\pi^2R^3$. Therefore the average energy density 
is given by 

$${{E_{dS}}\over  { 2\pi^2 R^3}}\;=\;
{ \Lambda \over {6\pi^2}}\;.\eqno(19)$$

\noindent Let us obtain now the distribution of gravitational
energy in the de Sitter space. In view of the spherical symmetry
we will be interested in calculating the density of energy per
unit radial distance $\varepsilon(r)$, which is obtained by
integrating ${1\over {8\pi}}\partial_r(eT^1)$ in $\theta$ and
$\phi$.  Thus $\varepsilon(r)$ yields the gravitational energy
contained between the spherical shells of radii $r$ and
$r\,+\,dr$. Considering $m=0$ in (15) we obtain upon integration
in the angular variables and differentiation in $r$:

$$ \varepsilon(r)\;=\;1\,+\,
{{2\beta^2-1}\over \sqrt{1-\beta^2}}\;.\eqno(20)$$

\noindent where we have set $\beta^2\,=\,{r^2\over R^2}$.
The function $\varepsilon(r)$ has the following properties.
In the range $0\le \beta \le 1\;\;$ $\varepsilon(r)$ vanishes
only for $r=0$. Moreover for $\beta=1$ it diverges:
$\varepsilon(\beta=1)\rightarrow \infty$. It is straightforward 
to check that this is a monotonically increasing
function, largely concentrated in the 
vicinity of $\beta=1$: $\varepsilon(\beta=0.1)=0.015\;;\;
\varepsilon(\beta=0.5)=0.423\;;\;\varepsilon(\beta=0.9)=
2,422$. The total energy contained inside the surfaces of 
radii $0.1R\;,\;0.5R\;,\;0.9R$ are given by
$E_g=5.01\times 10^{-4}R\;,\;E_g=0,067R\;,\;E_g=0.51R$,
respectively. 

Therefore almost half of the gravitational 
energy is located between $\beta=0.9$ and $\beta=1$.
This result is in total agreement with the phenomenological
features of the de Sitter solution, and can be verified in the
following way. The $g_{00}$ component of (11) can be written as

$$g_{00}\;=\;1\,+\,2\phi \;,$$

\noindent where $\phi$ is given by

$$ \phi\;=\; - {m\over r}\,-\,{1\over 6}\Lambda\,r^2\;.$$

\noindent $\phi$ is the potential in classical mechanics which
would induce  motion of a test particle approximately along
the geodesics of (11). Therefore even in the absence of a mass
$m$ a test particle would be subject to a radial acceleration

$$a\;=\;{1\over 3}\Lambda \,r\;,$$

\noindent directed away from $r=0$.

The acceleration increases with the distance $r$, indicating that
the gravitational field is more intense at points far from the
origin. Therefore when $m=0$ the gravitational given by (11)
acts on physical bodies as if there were a radially symmetric 
distribution of matter about the origin, beyond the cosmological
horizon, just as $m$ represents the mass of a black hole,
concentrated inside the black hole horizon.

This is precisely the picture we obtain from (20). 
By applying (9) to the de Sitter solution we find that 
the cosmological constant induces a spherically 
symmetric distribution of gravitational energy, concentrated
in regions distant from the origin, 
due to the gravitational field that acts on a test particle
placed in the vacuum de Sitter space. Such a field can be thought
as due to some matter distribution.

One may think of (11) as representing the gravitational field
of a spherical cavity inside some spherically symmetric
distribution of matter (this idea is discussed, 
for instance, in ref.\cite{ABS}). In this respect 
we recall that Einstein already conjectured  that de Sitter 
solution would correspond to a world with a surface
distribution of matter\cite{Pauli}. Such conjecture has found a
natural explanation within the present geometrical framework,
and shows that (9) yields a consistent expression for the
gravitational energy in  the de Sitter space.

We will briefly discuss how our procedure applies to the
anti-de Sitter solution. The latter is obtained by making
the replacement ${r^2\over R^2}\;\rightarrow 
-{r^2\over R^2}$ in (11). The anti-de Sitter space is a 
non-compact manifold with constant negative curvature.
The energy contained within a surface of constant radius $r_o$
can be easily calculated and reads

$$E_g\;=\;r_o \biggl(1-\sqrt {1+{{r_o^2}\over R^2}}\biggr)
\;,\eqno(21)$$

\noindent where we have ignored the mass term $m$. $\;\;r_o$ 
ranges from $0$ to $\infty$. Therefore as $r_o \rightarrow 
\infty$, we find that $E_g\rightarrow -\infty$. This is an
expected result, since the anti-de Sitter space is
non-compact. The density
of energy per unit radial distance $\varepsilon(r)$ in this
case is given by

$$\varepsilon(r)\;=\;1\,-\,{{1+{{2r^2}\over R^2}}\over 
\sqrt{1+{r^2\over R^2}}}\;.\eqno(22)$$

\noindent We find that $\varepsilon(r)=0$ only for $r=0$.
This point is also the only global maximum for $\varepsilon(r)$;
for $r\rightarrow \infty$ we clearly see that 
$\varepsilon(r) \rightarrow -\infty$. Thus  $\varepsilon(r)$
is a non-positive monotonically decreasing function.

\bigskip
\bigskip
\noindent {\bf IV. Discussion}\par

\noindent The definition of gravitational energy is a 
long-standing problem in the theory of general relativity.
Numerous attempts have been made in the past for a solution. 
This problem still attracts 
considerable attention in the literature, and
remains an important issue to be settled. Essentially all
of these previous attempts are in one or another way
unsatisfactory. In particular it is widely claimed that 
the gravitational energy cannot be localized. We do not share
this opinion. The mathematical structure of the TEGR shows 
that not only we do have a consistent and unambiguous 
definition of gravitational energy for asymptotically flat
spacetimes, naturally built in the  Hamiltonian formulation, 
but also that the gravitational energy is localized. The 
gravitational energy  in the framework of the TEGR is given by
expression (9). This expression has been successfully applied 
to a number of spacetimes, as we mentioned in section II, whose 
gravitational energy is already known. A
justification for the extension of this definition to more 
general spacetimes is not straightforward. In the case of 
asymptotically flat spacetimes the Hamiltonian constraint equation
can be written as $C=H-E_{ADM}=0$. We assume that this form of  
the constraint is a general feature of the theory, 
namely, that we can write
the Hamiltonian constraint as $C=H-E$ for an arbitrary spacetime,
since the constraint structure in general relativity
is fixed and does not depend on any particular topology. 

In the above we considered the de Sitter solution and concluded 
that the cosmological constant induces a distribution of 
gravitational energy largely concentrated in the vicinity of the
maximal spacelike radial distance $R$. 
This result is in total agreement with the fact that a
test particle in the de Sitter space is subject to a radial
acceleration directed away from the center of symmetry. Therefore
the outcome of our analysis complies with the 
phenomenological behaviour of a test particle in the de
Sitter space. To our knowledge this is the first time that such 
analysis has  been provided.

By inspecting equation (18) we see that if we make 
$\Lambda \rightarrow 0$ the total energy $E_g$ diverges. 
The vanishing of $\Lambda$ in (11) ammounts to a change from
a compact to a non-compact topology. Therefore such a change
is not smooth, as it requires an infinite ammount of energy.
This fact seems to indicate that, at the classical level, 
topology changing processes are forbidden.

\newpage

\noindent {\it Acknowledgements}\par
\noindent This work was supported in part by CNPQ.

\end{document}